\documentclass[a4paper,german,11pt,structabstract]{aa}

\usepackage{natbib}
\bibpunct{(}{)}{;}{a}{}{,}

\usepackage{multicol}
\usepackage{amsmath}
\usepackage{amssymb}
\usepackage{wasysym}
\usepackage{epsfig}
\usepackage{epsf}
\usepackage[dvips]{color}
\usepackage{graphicx,graphics}
\usepackage{float}
\usepackage{setspace,lineno}

\def\apjl{Astrophys. J. Letters}

\def\apjs{Astrophys. J. Suppl.}

\begin{document}
\begin{multicols}{1}

\title{The extrasolar planet GL 581 d: A potentially habitable planet?}
\titlerunning{GL 581 d habitable planet}

\author{P. von Paris\inst{1} \and S. Gebauer\inst{2} \and M. Godolt\inst{2} \and J.~L. Grenfell\inst{2}
\and P. Hedelt\inst{1}\thanks{now at:\newline CNRS, UMR 5804\newline
Laboratoire d'Astrophysique de Bordeaux\newline 2, rue de
l'Observatoire
\newline BP 89 - 33271 Floirac Cedex - France\newline
Universit\'e de Bordeaux\newline Observatoire Aquitain des Sciences
de l'Univers\newline 2 rue de l'Observatoire\newline BP 89 - 33271
Floirac Cedex, France } \and D. Kitzmann\inst{2} \and A.~B.~C.
Patzer\inst{2} \and H. Rauer\inst{1,2} \and B. Stracke\inst{1}}

\institute{Institut f\"{u}r Planetenforschung, Deutsches Zentrum
f\"{u}r Luft- und Raumfahrt, Rutherfordstr. 2, 12489 Berlin, Germany
\and Zentrum f\"{u}r Astronomie und Astrophysik, Technische
Universit\"{a}t Berlin, Hardenbergstr. 36, 10623 Berlin, Germany}

\abstract {}{The planetary system around the M star Gliese 581
contains at least three close-in potentially low-mass planets, GL
581 c, d, and e. In order to address the question of the
habitability of GL 581 d, we performed detailed atmospheric modeling
studies for several planetary scenarios.} {A 1D radiative-convective
model was used to calculate temperature and pressure profiles of
model atmospheres, assumed to be composed of molecular nitrogen,
water, and carbon dioxide. The model allows for changing surface
pressures caused by evaporation/condensation of water and carbon
dioxide. Furthermore, the treatment of the energy transport has been
improved in the model to account in particular for high CO$_2$,
high-pressure Super-Earth conditions.}
 {For four high-pressure scenarios of our study, the resulting
 surface temperatures were above 273 K, indicating a
potential habitability of the planet. These scenarios include three
CO$_2$-dominated atmospheres (95\% CO$_2$ concentration with 5, 10,
and 20 bar surface pressure) and a high-pressure CO$_2$-enriched
atmosphere (5\% CO$_2$ concentration with 20 bar surface pressure).
For all other considered scenarios, the calculated GL 581 d surface
temperatures were below the freezing point of water, suggesting that
GL 581 d would not be habitable then.

The results for our CO$_2$-dominated scenarios confirm very recent
model results by Wordsworth et al. (2010). However, our model
calculations imply that also atmospheres that are not
CO$_2$-dominated (i.e., 5\% vmr instead of 95\% vmr) could result in
habitable conditions for GL 581 d.}{}

\keywords{Astrobiology, Planets and satellites: atmospheres, Stars:
planetary systems, Stars: individual: Gliese 581, Planets and
satellites: individual: Gliese 581 d, }

\maketitle

\end{multicols}{1}

\section{Introduction}

Since the discovery of the first extrasolar planets around pulsars
\citep{wolszczan1992} and main-sequence stars \citep{mayor1995},
more than 400 planets and planet candidates have been detected, most
of which are Jupiter-mass objects. However, in recent years,
detections of smaller planets have been announced, and today 21
planets with (minimum) masses below 10 Earth masses ($m_{\oplus}$)
are known (e.g. \citealp{rivera2005}, \citealp{howard2009}). Two of
these planets, CoRoT-7 b \citep{leger2009} and GL 1214 b
\citep{charb2009}, have been discovered by the transit method and
produce a measurable radial velocity variation, hence their true
mass and mean density could be derived. They are the first two
examples of so-called "Super-Earths" (1 $m_{\oplus}$ $<m$ $\lesssim$
10 $m_{\oplus}$).

Modeling studies of the atmospheres of terrestrial (exo)planets
(e.g. \citealp{kasting1993}, \citealp{joshi1997},
\citealp{forget1997}, \citealp{pierre1998}, \citealp{selsis2002},
\citealp{Seg2003}, \citealp{Seg2005}, \citealp{Grenf2007pss},
\citealp{Grenf2007asbio}, \citealp{kitzmann2010}) have aimed at
characterizing the response of the atmospheric system to changes in
planetary (e.g. atmospheric composition, presence of clouds) and
stellar parameters (e.g. central star type, orbital distance). These
studies mostly assessed the potential surface habitability of such
planets and predicted their spectral appearance. Surface
habitability is commonly related to the presence of liquid water on
the surface, hence temperatures above 273 K, because liquid water
seems to be the fundamental requirement for life as we know it on
Earth.

In view of these modeling activities, the system Gliese 581 (GL 581)
with four planets (\citealp{bonfils2005}, \citealp{udry2007},
\citealp{mayor2009gliese}) is particularly interesting. It hosts at
least three potentially low-mass, therefore possibly terrestrial
planets. The one closest to the central star, GL 581 e, was
announced by \citet{mayor2009gliese} and has a minimum mass of 1.94
$m_{E}$ with an orbital distance of 0.03 AU to the star. The two
outer low-mass planets, GL 581 c and d, were discovered by
\citet{udry2007}. \citet{mayor2009gliese} refined the orbital
distances and minimum masses of these two planets, obtaining 5.36
$m_{E}$ (GL 581 c at a distance of 0.07 AU to the star) and 7.09
$m_{E}$ (GL 581 d at a distance of 0.22 AU to the star),
respectively.

Taking into account photometric (\citealp{lopez2006},
\citealp{mayor2009gliese}) and dynamical (\citealp{beust2008},
\citealp{mayor2009gliese}) constraints, the inclination of the GL
581 system most probably lies between 40$^{\circ}$ and 85$^{\circ}$.
This implies masses no higher than 1.56 times the minimum masses, so
all three low-mass planets are likely to be Super-Earths.

The habitability of GL 581 c and d has been investigated by
\citet{selsis2007gliese} and \citet{bloh2007}, based on the
discovery data published by \citet{udry2007}. In their study,
\citet{bloh2007} applied a box-model of the Earth, which
incorporates the carbonate-silicate cycle, requiring a surface
reservoir of liquid water and a tectonically active planet. It
includes the exchange of CO$_2$ between the mantle and crust of the
planet and its atmosphere by assuming parameterizations for
continental growth and spreading rate. The CO$_2$ partial pressure
and the stellar luminosity were related to the surface temperature
through a simple energy balance equation between stellar and emitted
thermal fluxes (see also \citealp{bloh2007pss} for more details).
\citet{selsis2007gliese}, on the other hand, used previous model
results for the habitable zone (HZ) from the seminal study of
\citet{kasting1993} to estimate the boundaries of the HZ in the GL
581 system. The results of \citet{kasting1993} were obtained with a
1D radiative-convective model. Furthermore, \citet{selsis2007gliese}
used model results for early Mars from \citet{mischna2000} to
illustrate the uncertainty in the outer limit of the HZ because of
the (possible) presence of CO$_2$ clouds.

Both \citet{selsis2007gliese} and \citet{bloh2007} concluded that
the inner planet GL 581 c is unlikely to be habitable, because it is
closer to the star than the inner boundary of the HZ, whereas the
outer planet, GL 581 d, just might be habitable. Based on the
calculations by \citet{selsis2007gliese}, \citet{mayor2009gliese}
concluded that GL 581 d lies in the HZ of its central star,
considering that the refined orbit means it receives more than 30 \%
more stellar energy than previously thought.

Very recently, \citet{wordsworth2010} presented 1D modeling studies
of high CO$_2$ atmospheres of GL 581 d, varying the CO$_2$ pressure
and other parameters such as relative humidity and surface albedo.
They found habitable surface conditions with CO$_2$ partial
pressures of $\geq$5 bar. In a similar approach to
\citet{wordsworth2010}, we applied a different 1D
radiative-convective model to the atmosphere of GL 581 d considering
different appropriate planetary scenarios. By calculating
atmospheric pressure, temperature, and water profiles, a reasonable
range of planetary surface conditions has been investigated with
regard to potential habitability.

The paper is organized as follows: Sect. \ref{thesystem} describes
the properties of the planetary system GL 581. The model used is
described in Sect. \ref{model}. A description of the model input
parameters and of the runs performed is given in Sect.
\ref{modinput}. Results are described and discussed in Sect.
\ref{resultsect}. We give our conclusions in Sect. \ref{concl}.


\section{The planetary system GL 581}

\label{thesystem}

\subsection{Main properties of the star GL 581}

GL 581 is a quiet M3 star \citep{udry2007}. Its main parameters are
summarized in Table \ref{starpar}.

\begin{table}[H]

\caption{Main properties of GL 581}

\label{starpar}

\begin{center}
\resizebox{\hsize}{!}{\begin{tabular}{lll}

  \hline
  Property            &  Value              & Reference           \\
  \hline
  Type                &  M3                 &\citet{udry2007}     \\
  Mass                &  0.31 M$_{\odot}$    &\citet{bonfils2005}  \\
  Luminosity          &  0.013 L$_{\odot}$   &\citet{udry2007}, \citet{bonfils2005}\\
                      &  0.0135 L$_{\odot}$  &\citet{selsis2007gliese}\\
  Radius              &  0.38 R$_{\odot}$    &\citet{lacy1977}, \citet{johnson1983}\\
                      &  0.29 R$_{\odot}$    & \citet{chabrier2000}\\
  T$_{\mathrm{eff}}$  &  3190 K            &calculated with the Stefan-Boltzmann law\\
                      &  3249 K            & \citet{johnson1983}\\
                      &  3760 K            & \citet{butler2006}\\
  $[$Fe/H$]$          &  -0.10              & \citet{johnson2009}\\
                      &  -0.25              & \citet{bonfils2005}\\
  distance            &  6.27 pc            & \citet{butler2006}\\
                      &  6.53 pc           & \citet{johnson1983}, \citet{lacy1977}\\

\end{tabular}}
\end{center}

\end{table}

\begin{figure}[H]
\resizebox{\hsize}{!}{
\includegraphics[width=350pt]{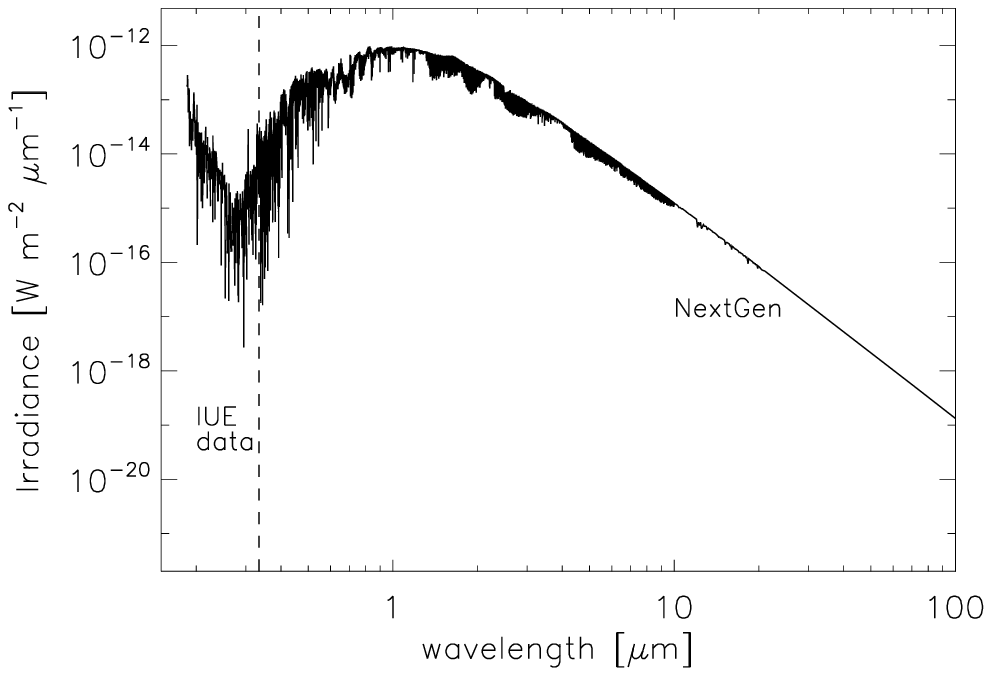}}\\
  \caption[Derived high-resolution spectrum of GL 581]
  {Derived high-resolution spectrum of GL 581 (data sources as indicated).}\label{gliese_highres}
\end{figure}

In order to perform proper atmospheric modeling studies, we need a
spectrum of GL 581. Therefore, we derived a high-resolution spectrum
(Fig. \ref{gliese_highres}) from two sources, i.e. an UV spectrum
measured by the IUE (International Ultraviolet Explorer) satellite
in 1987 and a synthetic Nextgen model spectrum
\citep{hauschildt1999}.

The model parameters of the available NextGen model spectrum used
here are $T_{\mathrm{eff}}$=3200 K, $\log$ g= 4.5, $[$Fe/H$]$=0.0
and $R_{\mathrm{GL581}}$=0.38 $R_{\odot}$. The synthetic spectrum
was calculated for wavelengths from 0.334-971 $\mu$m. It was merged
with the IUE spectrum from 0.185-0.334 $\mu$m.

Based on the 6.27 pc distance of GL 581 \citep{butler2006}, the
spectrum was scaled to the orbital distance of GL 581 d (0.22 AU,
\citealp{mayor2009gliese})  and binned to the spectral intervals
required for the model code (see Sect. \ref{model}).

The orbit of GL 581 d could be highly eccentric (Table
\ref{planpar}). The mean stellar flux $\overline{F}$ received over
an eccentric orbit is given by

\begin{equation}\label{meanflux}
\overline{F}=\frac{F(a)}{(1-e^2)^{0.5}}
\end{equation}

where $F(a)$ is the flux received by the planet at the distance of
its semi-major axis $a$. As shown by \citet{williams2002} with a 3D
model approach, the overall climate of a planet on an eccentric
orbit behaves almost as if the planet were constantly receiving this
averaged flux $\overline{F}$. For GL 581 d, the application of Eq.
(\ref{meanflux}) results in an increase of the stellar flux of about
8\% compared to the circular case.

Assuming that the atmospheric response timescales to changes in
stellar flux (which are on the order of weeks/months) are comparable
to the orbital period of the planet (about teo months),
$\overline{F}$ was used for the model calculations. Note, though,
that this approach depends on the properties of the central star and
the planetary system. For larger orbital distances, for instance,
flux variations caused by orbital eccentricities should be taken
into account.

\subsection{Properties of the planet GL 581 d}

The semi-major axis is known from the measured orbital periods and
the mass of the central star via Kepler's third law. This we took
from \citet{mayor2009gliese} to be $a_{\mathrm{GL581d}}=0.22$ AU.
The best fit of the radial velocity data was obtained for a highly
eccentric orbit with $e_d$=0.38 (see \citealp{mayor2009gliese}).

In order to be consistent with the studies of
\citet{selsis2007gliese} and \citet{bloh2007}, the stated minimum
mass was assumed to be the true mass $m$ of the planet. The
planetary radius $r$ is then taken from a theoretical mass-radius
relationship for terrestrial planets \citep{sotin2007}, finally
yielding the surface gravity.

Usually, planets in the HZ of M stars (or closer to the star) are
assumed to be tidally locked in a synchronous rotation, i.e. they
rotate with the same rate $\omega$ as their orbital period $P$ (e.g.
\citealp{lammer2007a}, \citealp{scalo2007}). However, as already
noted by \citet{goldreich1968}, \citet{griess2005},
\citet{levrard2007} and \citet{correia2008}, planets that show
significant eccentricities -- possibly like GL 581 d -- are unlikely
to become locked in a 1:1-resonance. Therefore, GL 581 d is not
considered to rotate synchronously in this study.

An important model parameter for 1D radiative-convective models is
the surface albedo. One approach used in these models is to adjust
the surface albedo in the model (e.g., \citealp{kasting1993},
\citealp{Grenf2007pss}, \citealp{goldblatt2009faintyoungsun}) so
that the models reproduce prescribed reference scenarios (e.g.
modern Earth with a surface temperature of 288 K). Therefore, the
values of the model surface albedos largely depend on the impact of
clouds on the surface temperature. This impact, however, is unknown
for Super-Earths. Hence, instead of an arbitrary choice of the
surface albedo used in the tuning of 1D models, we adopted the
measured surface albedo of Earth here. Thus, the unconstrained
effect of clouds in Super-Earth atmospheres is explicitly excluded
from our cloud-free simulations. The surface albedo value used here
is $A_{\rm{surf}}$=0.13 (Earth mean, see \citealp{kitzmann2010} and
\citealp{rossow1999}).

Table \ref{planpar} summarizes the assumed planetary parameters of
GL 581 d.

\begin{table}[H]

\caption{Planetary parameters of GL 581 d. }

\label{planpar}

\begin{center}
\resizebox{\hsize}{!}{\begin{tabular}{lrrrrc}

  \hline
  Property            &  Value              & Reference\\
  \hline
  Mass                &  7.09 m$_{\oplus}$   &\citet{mayor2009gliese}\\
  Orbital distance    &  0.22 AU            &\citet{mayor2009gliese}\\
  Eccentricity        &  0.38               &\citet{mayor2009gliese}\\
  Radius              &  1.71 r$_{\oplus}$   & after \citet{sotin2007}\\
  Gravity             &  23.76 ms$^{-2}$    & calculated\\
  Surface albedo      &  0.13         & Earth mean, \citet{rossow1999}\\

\end{tabular}}
\end{center}

\end{table}

\section{Atmospheric model}

\label{model}

\subsection{General model description}

A 1D, cloud-free radiative-convective column model was used for the
calculation of the atmospheric structure, i.e. the temperature and
pressure profiles for different GL 581 d scenarios.

The model is originally based on the one described by
\citet{kasting1984water} and \citet{kasting1984}. Further
developments are described by e.g. \citet{kasting1988},
\citet{kasting1991}, \citet{kasting1993}, \citet{mischna2000} and
\citet{pavlov2000}. Additional updates of the thermal radiation
scheme of the model have been introduced by \citet{Seg2003}. The
model version used here is based on the version of
\citet{vparis2008} where a more detailed model description is given.
The current model uses the radiative transfer scheme MRAC in the IR
(see the introduction of this scheme in \citealp{vparis2008}). The
water profile in the model is calculated based on the relative
humidity distribution of \citet{manabewetherald1967}. Above the cold
trap, the water profile is set to an isoprofile at the cold trap
value.

The model considers N$_2$, H$_2$O, and CO$_2$ as atmospheric gases.
Other radiative trace species might be present in the atmospheres of
exoplanets (e.g. SO$_2$, O$_3$, or CH$_4$), which could alter the
radiative budget significantly. But because the presence of these
gases highly depends on the planetary scenario (e.g., outgassing,
volcanism, formation, biosphere), our model atmospheres are
restricted to the two most important greenhouse gases of the Earth's
atmosphere (H$_2$O and CO$_2$), using N$_2$ as additional background
gas. This is based on the observation that in the solar system all
terrestrial atmospheres (Venus, Earth, Mars, Titan) contain
significant amounts of N$_2$.

Temperature profiles from the surface up to the mid-mesosphere are
calculated by solving the equation of radiative transfer and
performing convective adjustment, if necessary. The convective lapse
rate is assumed to be adiabatic. Water profiles are calculated by
assuming a fixed relative humidity profile
\citep{manabewetherald1967} through the troposphere.

As a globally averaged 1D column model, the model does not
incorporate dynamical processes such as winds and latitudinal energy
transport. Thus potentially important 3D effects are not included.
For example, the effect of atmospheric collapse and global energy
re-distribution on tidally locked planets was investigated by
\citet{joshi1997} and \citet{joshi2003}. The study of
\citet{spiegel2008} showed that slowly rotating planets feature a
very efficient energy redistribution because of the weaker Coriolis
force (compared to Earth). Hence a 1D model could give a reasonably
accurate picture of global mean temperatures (and therefore,
habitability) on planets orbiting M stars. In addition to rotation,
also atmospheric optical depth and mass of a planet are important
for assessing temperature gradients \citep{joshi1997}. Therefore, 3D
atmospheric modeling of GL 581 d is needed in the future to
investigate its potential habitability in more detail.

\subsection{Computational details}
\label{details}

In this work, we employed an improved version of the model of
\citet{vparis2008}.

\begin{enumerate}
  \item

  In the present study, CO$_2$ condensation is included for the
calculation of the adiabatic lapse rate in the convective regime
according to the treatment by \citet{kasting1991} and
\citet{kasting1993}. The saturation vapor pressure of CO$_2$ is
taken from \citet{Ambrose1956}. CO$_2$ condensation is generally
assumed to occur when the atmosphere is (super-)saturated with
respect to CO$_2$. This criterion is expressed by the super
saturation ratio $S_{\rm{sat}}$, which should be higher than unity:

\begin{equation}\label{co2glan}
    \frac{p_{\rm{CO_2}}}{P_{\rm{sat},CO_2}} = S_{\rm{sat}}\geq 1
\end{equation}

where $p_{\rm{CO_2}}$ is the partial CO$_2$ pressure and
$P_{\rm{sat},CO_2}$ the saturation vapor pressure of CO$_2$. The
numerical value of $S_{\rm{sat}}$ used here is taken to be 1.34,
based on measurements by \citet{glandorf2002}. This implies that the
atmosphere has super-saturated with respect to CO$_2$ for CO$_2$
condensation to occur. Note that \citet{kasting1991} and
\citet{kasting1993} assumed $S_{\rm{sat}}$=1, i.e. the model
atmosphere was assumed to be saturated.

\item

The calculations now also include Rayleigh scattering by water vapor
($\sigma_{\rm{ray,H_2O}}$, in [cm$^{2}$]) calculated via

\begin{equation}\label{rayleighwater}
\sigma _{\rm{ray,H_2O}}(\lambda)=4.577\cdot 10^{-21} \cdot
\left(\frac{6+3\cdot D}{6-7\cdot D}\right)\frac{r^2}{\lambda^4}
\end{equation}

where $D$ is the depolarization ratio, $r$ the refractivity and
$\lambda$ the wavelength in $\mu$m. The numerical factor 4.577$\cdot
10^{-21}$ is taken from \citet{Allen1973}.

This work assumes $D=0.17$ from \citet{marshall1990}. The
refractivity is calculated as $r=0.85 \cdot r_{\rm{dry air}}$
\citep{edlen1966}. The refractivity of dry air is obtained from an
approximate formula given by \citet{bucholtz1995}.

\item
The heat capacity of water vapor has been included in the
calculations based on the so-called Shomate equation
\citep{parks1940} with parameters from \citet{chase1998}.

\item

One critical issue for the IR radiative transfer scheme used here is
the so-called binary species parameter $\eta$ (\citealp{Mlawer1997},
\citealp{Cola2003}, \citealp{vparis2008}). In previous studies, the
variation of absorption cross sections of different mixtures was
assumed to be linear according to the relative concentrations of the
major absorbers (here, H$_2$O and CO$_2$). For atmospheres where the
relative concentrations of water and carbon dioxide vary by several
orders of magnitude, this approach depends on the assumed reference
ratio. Hence, a logarithmic $\eta$ grid has been introduced where
relative concentrations are allowed to vary by 15 orders of
magnitude.

\item

The IR H$_2$O continuum calculation used in \citet{vparis2008} has
been replaced by the CKD continuum formulation
(\citealp{clough1989}, \citealp{schreier2003}). In this way, the
continuum is no longer restricted to the atmospheric window region
(between 8 and 12 $\mu$m) and incorporates contributions from H$_2$O
self as well as foreign continua and, for the first time in the
context of exoplanet 1D models, the CO$_2$ foreign continuum in the
IR.

  \item
The temperature-pressure grid for the absorption coefficient
calculations in the IR has been extended to $T$=100 K and $p$=1000
bar. This was done as described by \citet{vparis2008}, using line
parameters from \citet{rothman1995}.

  \item

The model has been further modified to account for changing pressure
levels during the calculations. Under some conditions described
here, the atmospheric content of water or carbon dioxide is
controlled by the surface temperature caused by evaporation or
condensation. The surface pressure is thus determined by the surface
temperature and must be re-calculated after each model time step to
adjust to the newly calculated value of the surface temperature. The
model pressure grid is therefore no longer fixed, but dynamically
determined between two consecutive iterations in the model.

\end{enumerate}

The IR radiative transfer updates mentioned above were tested
against high-resolution line-by-line (lbl) calculations with the
MIRART radiative transfer code \citep{schreier2003}, which was found
to compare well with other lbl codes (e.g.,
\citealp{melsheimer2005}).

The total IR flux was within 5\% of the lbl values for all
atmospheric levels. Fluxes for each spectral interval of the
radiative transfer code were usually within 10\% of the lbl values,
except for a few near-IR bands in the middle to upper atmosphere.
However, these bands do not contribute strongly to the cooling rate
(less than 10$^{-6}$ ) or the stratospheric energy budget, hence
these deviations are insignificant for the calculation of the
temperature profiles. Further comparative studies with the improved
model are presented in Appendix \ref{comparison}.

\section{Model input for GL 581 and performed runs}

\label{modinput}

Important properties characterizing planetary atmospheres are the
total surface pressure and the composition of the atmosphere. These
determine the greenhouse effect and the volatile reservoir, hence
the habitability of a planet. Atmospheric pressure and composition
are determined by the accretion and outgassing history, whether from
the interior or from impacts  (e.g. \citealp{pepin1991}),
atmospheric loss because of escape to space (e.g.
\citealp{kulikov2007}, \citealp{kulikov2006}) or incorporation into
planetary surface material. But these processes are not known for
the GL 581 system. Hence, in order to study the influence of
atmospheric properties on the potential habitability of GL 581 d, we
performed a parameter study considering different atmospheric
scenarios.

\begin{figure*}
  \centering
\begin{center}\includegraphics[width=270pt]{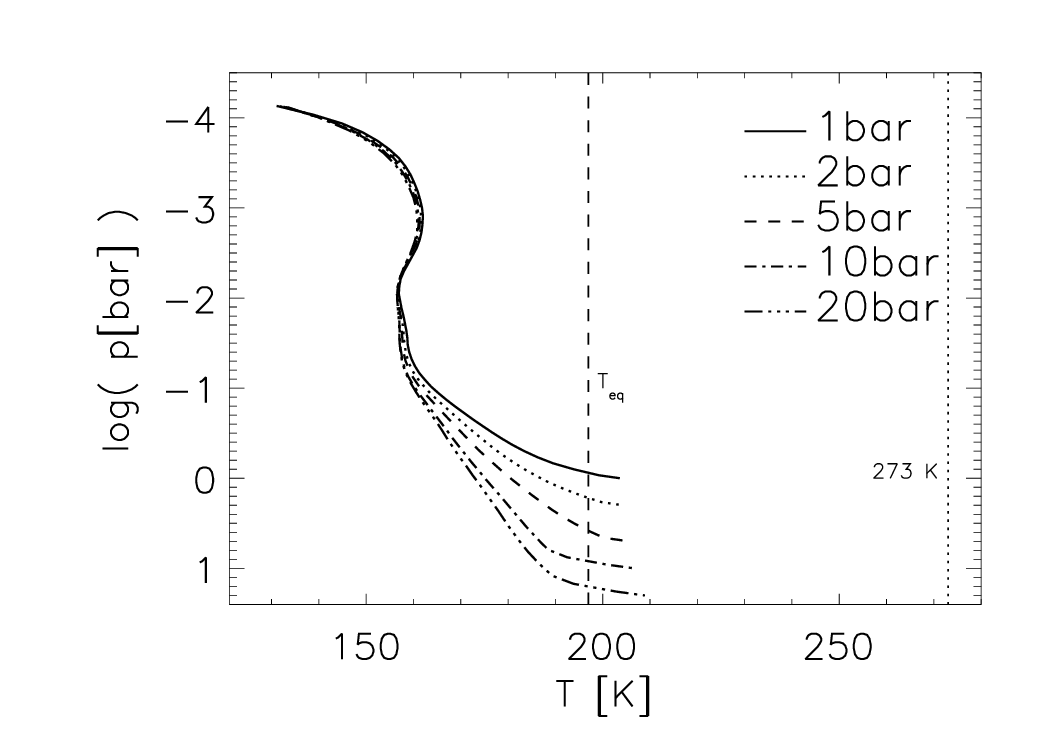}\\\end{center}
  \caption[Temperature-pressure profiles for the low CO$_2$ runs]
  {Temperature-pressure profiles for the low CO$_2$ (355 ppm CO$_2$) runs of Table \ref{listofruns}.
  The equilibrium temperature of the planet (dashed) and melting temperature of water (dotted) are indicated as vertical lines. }
  \label{temperature_low}
\end{figure*}

We assumed that the initial water delivered to the planet via
outgassing and impacts has been retained, hence a reservoir of
liquid water is present on the planetary surface. This assumption
was also made by \citet{selsis2007gliese}, \citet{bloh2007} and
\citet{wordsworth2010} for the GL 581 planetary system.

The total surface pressure (1, 2, 5, 10, 20 bar) and CO$_2$
concentration (0.95, 0.05 and 355 ppm vmr, respectively) were
varied. N$_2$ is assumed to act as a filling background gas. The
range of surface pressures was chosen to represent scenarios adopted
in the literature for early Earth and early Mars in terms of
atmospheric column density. The values of CO$_2$ concentrations are
chosen to represent modern Earth (355 ppm CO$_2$ concentration),
early Earth (5\% CO$_2$) and Mars or Venus (95\% CO$_2$) conditions,
i.e. representative values of terrestrial atmospheres within the
solar system. Table \ref{listofruns} summarizes the set of runs for
GL 581 d performed here.

\begin{table}[H]
  \caption{Atmospheric scenarios for GL 581 d
  }\label{listofruns}
\begin{center}
\resizebox{\hsize}{!}{\begin{tabular}{lcc}
 \hline
   Set  &   $p$ [bar]     & CO$_2$ vmr           \\
  \hline
    G1 (low CO$_2$)  & 1,2,5,10,20       &3.55 $\cdot$ 10$^{-4}$           \\
    G2 (medium CO$_2$) & 1,2,5,10,20       &0.05                              \\
    G3  (high CO$_2$)& 1,2,5,10,20       &0.95                       \\
\end{tabular}}
\end{center}

\end{table}

\begin{figure*}
\begin{center}\includegraphics[width=270pt]{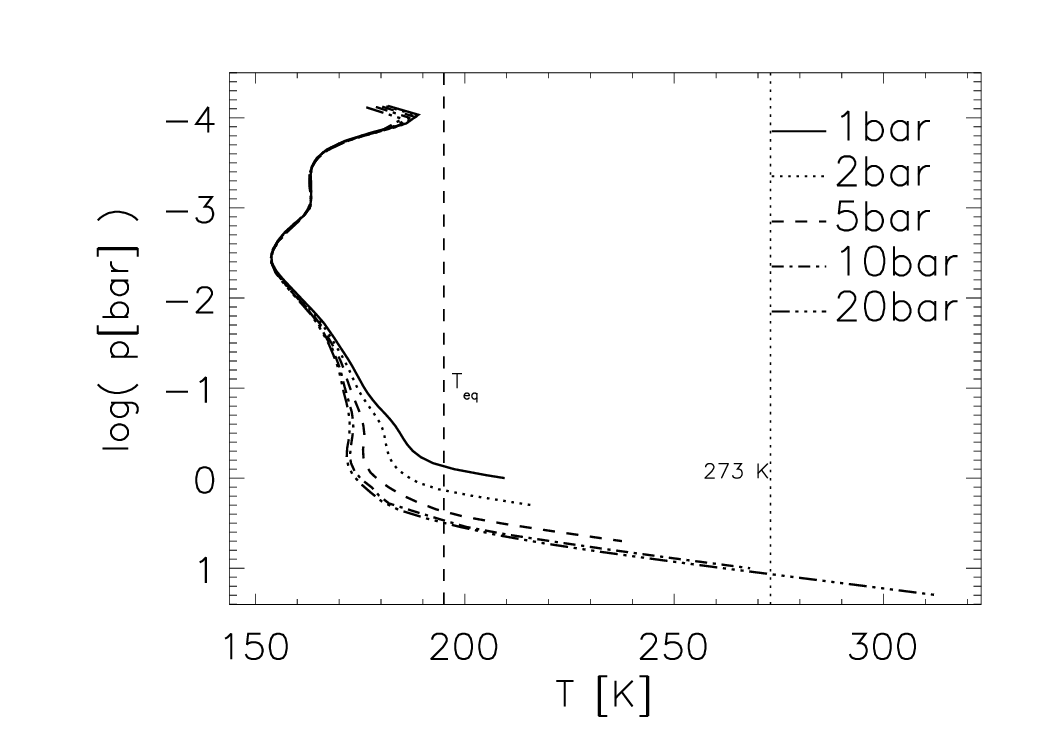}\\\end{center}
  \caption[Temperature-pressure profiles for the medium CO$_2$ runs]
  {Temperature-pressure profiles for the medium CO$_2$ (5 \% CO$_2$) runs of Table \ref{listofruns}.
  The equilibrium temperature of the planet (dashed) and freezing temperature of water (dotted) are indicated as vertical lines.}
  \label{temperature_medium}
\end{figure*}

\section{Results and discussion}

\label{resultsect}

\subsection{Temperature profiles}

Figure \ref{temperature_low} shows the temperature-pressure profiles
for the low CO$_2$ set of runs (G1) of Table \ref{listofruns}. For
all runs the surface temperatures are far lower than the freezing
point of water (273 K, indicated by the dotted vertical line in Fig.
\ref{temperature_low}), which implies that liquid water is not
present on the surface of the planet. Upon increasing the surface
pressure from 1 bar to 20 bar, the surface temperatures increase by
4.2 K (from 203.6 K to 208.9 K).

\begin{figure*}
\begin{center}\includegraphics[width=270pt]{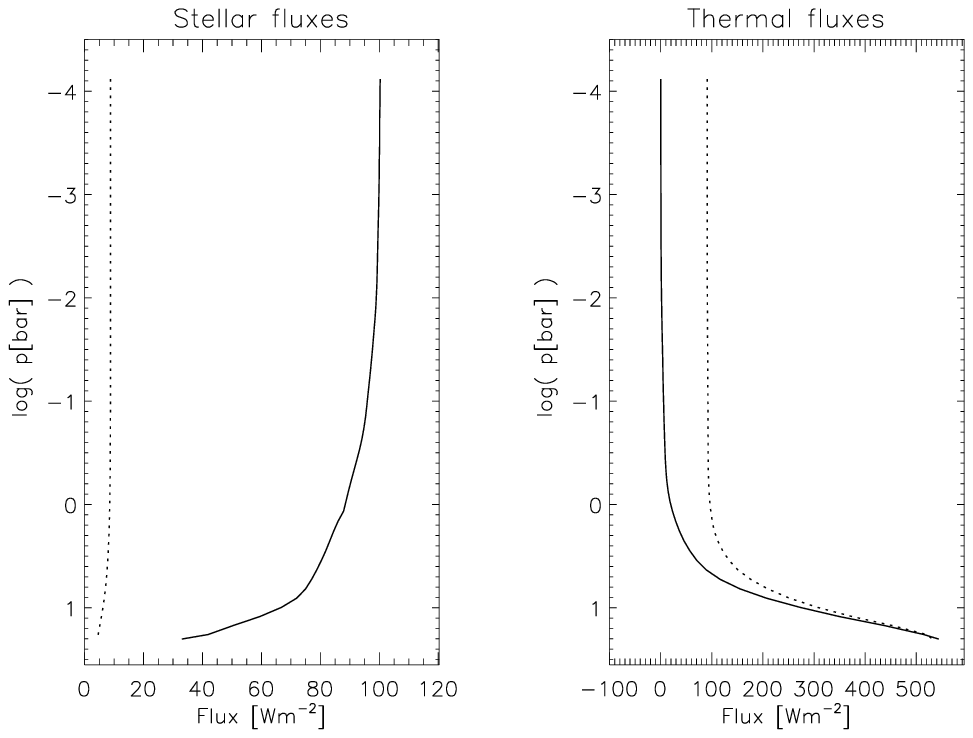}\\\end{center}
  \caption[Net fluxes of the 20 bar run of set G2 (medium CO$_2$)]
  {Net fluxes of the 20 bar run of set G2 (medium CO$_2$): Stellar (left) and thermal (right). Downwelling (solid) and
  upwelling (dotted) fluxes are shown.}
  \label{netfluxes_medium}
\end{figure*}

Also shown in Fig. \ref{temperature_low} is the equilibrium
temperature $T_{\rm{eq}}$ of the planet. It is about 197 K, hence
habitable conditions would require a greenhouse effect (GHE) of 76
K. As can be inferred from Fig. \ref{temperature_low}, the actual
GHE in the model atmospheres is about 6 to 12 K, which is much less
than the approximately 30 K GHE provided by the modern Earth
atmosphere.

An interesting feature of the temperature profiles of the low CO$_2$
case is the absence of a convective troposphere for all G1 runs.
These atmospheres are all in radiative equilibrium.

Figure \ref{temperature_medium} shows the temperature-pressure
profiles for the medium CO$_2$ set of runs (G2) of Table
\ref{listofruns}. For the 20 bar run, the value of the surface
temperature was 313.3 K, hence exceeded the freezing point of water.
Thus, in this particular scenario GL 581 d might be habitable. The
other runs with 1, 2, 5, and 10 bar though all showed surface
temperatures below 273 K. Note, however, that for the 10 bar run the
obtained surface temperature is 268.1 K, which is fairly close to
the freezing point of water.

In contrast to the low CO$_2$ case, the increase of surface pressure
from 1 to 20 bar has a huge effect on surface temperature, which
increased by about 105 K. This is caused by an enhanced absorption
of stellar radiation by CO$_2$ and H$_2$O near-IR absorption bands
and a massive greenhouse effect, which results in the onset of
convection for runs with higher surface pressures.

Figure \ref{netfluxes_medium} illustrates this effect, showing net
(i.e., frequency-integrated) stellar and thermal downward ($F_d$)
and upward ($F_u$) fluxes for the medium CO$_2$ (G2) 20 bar run.
Much of the incoming stellar radiation ($\sim$ 70 \%) is absorbed by
CO$_2$ and water in the lower atmosphere, as illustrated by the left
panel in Fig. \ref{netfluxes_medium}. The difference $C_r=F_u-F_d$
for the thermal fluxes is the radiative cooling. A strong GHE is
indicated by a low value of $C_r$. In the lower atmosphere of the 20
bar run, both thermal components balance each other, as can be seen
in Fig. \ref{netfluxes_medium}, meaning that the thermal radiation
is efficiently trapped in the atmosphere. The value of $C_r$ in the
bottom atmosphere layer is about 5 Wm$^{-2}$, which corresponds to
about 1 \% of the surface emission ($\sigma_B$ $T_{\rm{surf}}^4$
$\approx$ 540 Wm$^{-2}$, $\sigma_B$ Stefan's constant,
$T_{\rm{surf}}$=313 K surface temperature). On Earth, the value of
$C_r$ is about 70 Wm$^{-2}$, about 20 \% of the surface emission.

The atmospheric structure of the medium CO$_2$ runs in Fig.
\ref{temperature_medium} is very different from the low CO$_2$ runs
in Fig. \ref{temperature_low}. Firstly, the temperature inversion in
the upper atmosphere of the G2 runs at pressures below 10 mbar is
much more pronounced ($\sim$ 30 K) than for the low CO$_2$ runs.
This is owing to the strong absorption of stellar radiation by
CO$_2$ and H$_2$O in the near-IR bands (at 2, 2.7 and 4.3 $\mu$m).

\begin{figure*}
\begin{center}\includegraphics[width=270pt]{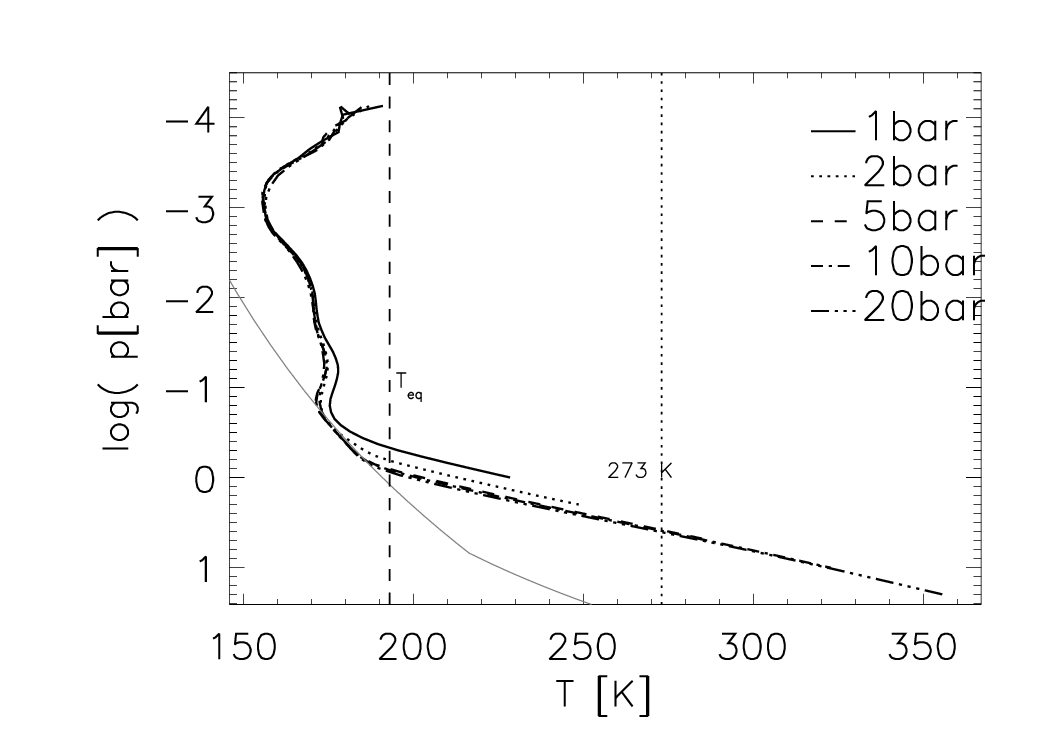}\\\end{center}
  \caption[Temperature-pressure profiles for set G3]
  {Temperature-pressure profiles for set G3 of Table \ref{listofruns} (95\% CO$_2$).
  The CO$_2$ condensation curve (with $S_{\rm{sat}}$=1.34 in Eq.
\ref{co2glan}) is shown in light gray.}
  \label{temperature_high}
\end{figure*}

Secondly, convective tropospheres develop for the 5, 10, and 20 bar
runs. These extend to an altitude of about 0.5-1 surface scale
heights (about 2-5 bar) above the surface, which is comparable to
the troposphere extension on present Earth. Hence, the lapse rate is
much steeper in the medium CO$_2$ cases than in the low CO$_2$
cases. For example, the convective lapse rate in the 20 bar medium
CO$_2$ case is about 22 K km$^{-1}$ near the surface (the dry
adiabatic lapse rate varies linearly with gravity), whereas the
radiative lapse rate for the 20 bar low CO$_2$ run is only about 9 K
km$^{-1}$. The value of 22 K km$^{-1}$ is very close to the dry
adiabatic lapse rate of 23 K km$^{-1}$ in the medium CO$_2$ case.
Despite the high surface temperature of 313 K and a corresponding
partial pressure of water of about 70 mbar, the water concentrations
are only of the order on 10$^{-3}$ vmr near the surface. Hence, the
lapse rate is close to the dry adiabat (see discussion in
\citealp{ingersoll1969}), even if appreciable amounts of water are
present in the atmosphere.

In Fig. \ref{temperature_high}, the temperature-pressure profiles
for the high CO$_2$ set of runs (G3) are shown. Except for the
low-pressure runs with 1 and 2 bar surface pressure, all scenarios
showed surface temperatures above 273 K. Hence, the results suggest
that CO$_2$-rich atmospheres may imply habitable conditions on GL
581 d.

The atmospheric structure in the high-CO$_2$ case is different from
those in the low and medium CO$_2$ cases. Even the 1 and 2 bar runs
now show convective tropospheres, albeit not very extended ones.
More massive tropospheres develop in the 5 bar, 10 bar, and 20 bar
runs owing to the onset of CO$_2$ condensation in the middle
atmosphere, as indicated by the CO$_2$ condensation curve in Fig.
\ref{temperature_high}. The respective tropopauses are located 3-5
surface scale heights above the planetary surface. On present Earth,
this would correspond to tropopause heights of about 20-40 km
compared to the roughly 10 km tropopause altitude observed today.

The tropospheres in the 5, 10, and 20 bar high CO$_2$ runs are
divided into two regimes, an upper troposphere with CO$_2$
condensation (see CO$_2$ condensation curve in Fig.
\ref{temperature_high}) and a lower troposphere with H$_2$O
condensation. This temperature structure is comparable to
atmospheric structures calculated for models of the early Mars
atmosphere (\citealp{kasting1991}, \citealp{mischna2000},
\citealp{Cola2003}).

\begin{figure*}
\begin{center}\includegraphics[width=270pt]{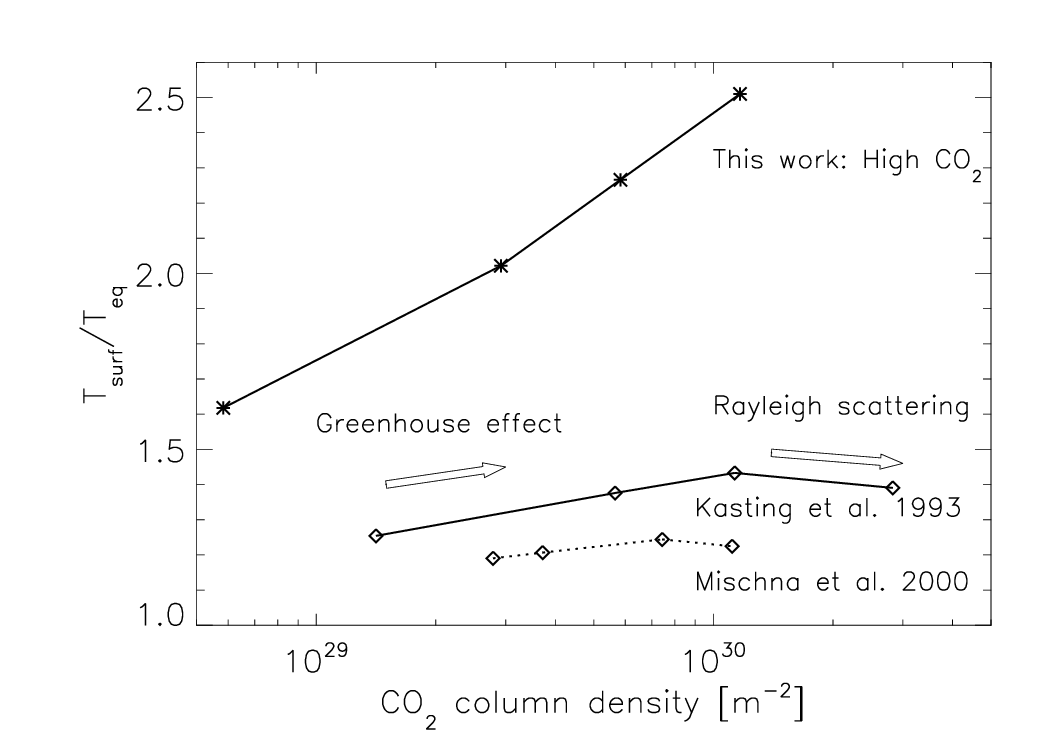}\\\end{center}
  \caption[Maximum greenhouse effect: Response of surface temperature to increase of atmospheric CO$_2$ content.]
  {Maximum greenhouse effect: Response of surface temperature to increase of atmospheric CO$_2$ content.}\label{maxghe}
\end{figure*}

From Fig. \ref{temperature_high} it is also evident that temperature
profiles are similar to each other for the 5, 10, and 20 bar runs.
The temperature profiles of these runs differ by about 2-5 K at
equal pressures, as is the case for the 10 and 20 bar medium CO$_2$
runs (Fig. \ref{temperature_medium}). This is because of two
reasons. Firstly, the lower atmospheres of the 5, 10, and 20 bar
high-CO$_2$ runs become optically thick for thermal radiation, as is
observed for the medium CO$_2$ 10 and 20 bar runs. Secondly, an
increase in surface pressure does not result in an increase in
planetary albedo because GL 581 emits only a negligible amount of
radiation in the spectral range where the Rayleigh cross section is
largest. Therefore, the global energy balance is no longer affected
by the lower part of the model atmosphere, hence temperature
profiles do not differ significantly.

A widely used assumption for the assessment of the HZ is the
so-called maximum greenhouse, introduced by \citet{kasting1993} for
an Earth-like planet around the Sun. With increasing amounts of
CO$_2$ in the atmosphere of a planet located near the outer boundary
of the HZ, the GHE will become more and more saturated, i.e. the
optical depth is near or higher than unity for all CO$_2$ bands.
Then, any further increase of CO$_2$ will only increase Rayleigh
scattering, hence increase the planetary albedo. Thus, for
increasing CO$_2$ partial pressure, surface temperatures will show a
maximum and then decrease with increasing CO$_2$ pressure. The same
behavior was found for early Mars \citep{mischna2000}. The different
behavior for our high CO$_2$ runs compared to the results of
\citet{kasting1993} and \citet{mischna2000} is illustrated in Fig.
\ref{maxghe}. They are summarized in terms of atmospheric column
density and normalized surface temperature
$T_{\rm{surf}}/T_{\rm{eq}}$ where $T_{\rm{surf}}$ is the surface
temperature of the planet. As is obvious from this figure, the
high-CO$_2$ runs from this work do not feature a maximum greenhouse
effect. However, a maximum greenhouse effect will probably occur at
still higher pressures, hence column densities (not investigated
here). The absence of a maximum greenhouse effect for the runs of
this work is caused by four factors.

Firstly, GL 581 emits much more radiation in the near- to mid-IR,
and less in the visible than the Sun. Consequently, the contribution
of Rayleigh scattering to the planetary albedo is much less
efficient for planets around GL 581 than around the Sun because of
the $\lambda^{-4}$-dependence of the Rayleigh scattering cross
section. Secondly, the stronger near-IR emission of GL 581 leads to
more heating by near-IR absorption bands of H$_2$O and CO$_2$.
Thirdly, the simulations of \citet{kasting1993} were done at
constant surface temperatures of 273 K (hence, constant partial
pressure of 6.5 mbar water), which neglects the positive feedback
provided by increased water vapor at higher surface temperatures.
Fourthly, because of the higher gravity of GL 581 d compared to
Earth, the same column amount of CO$_2$ (i.e., x coordinate in Fig.
\ref{maxghe}) is reached at much higher pressures, e.g. 20 bar on GL
581 d compared to 8 bar on Earth. The pressure broadening of
absorption lines then leads to an enhanced absorption in the line
wings. Because the line centers are usually optically thick, a
higher absorption coefficient in the line wings can significantly
increase the overall absorption of radiation. It was recently
suggested that this kind of behavior could have helped to warm the
early Earth by invoking higher N$_2$ partial pressures than today
\citep{goldblatt2009faintyoungsun}. It is also mainly responsible
for the fact that surface temperatures for the 1 bar high CO$_2$
case are about 85 K lower than for the 20 bar medium CO$_2$ case,
although both atmospheres contain the same amount of CO$_2$. When
interpreting this result, one has to bear in mind the potential
uncertainties related to the radiative transfer in high-pressure
atmospheres (see sensitivity study in Appendix \ref{sensstudy}).

\subsection{Comparison with other studies of GL 581 d}

For several model scenarios (5, 10, and 20 bar high CO$_2$ and 20
bar medium CO$_2$ runs) of this work, surface temperatures were
found to be above 273 K, i.e. these results imply habitable surface
conditions on GL 581 d. In all other scenarios, GL 581 d was found
to be uninhabitable with surface temperatures below 273 K.

\begin{figure}[H]
 \begin{center}  \includegraphics[width=250pt]{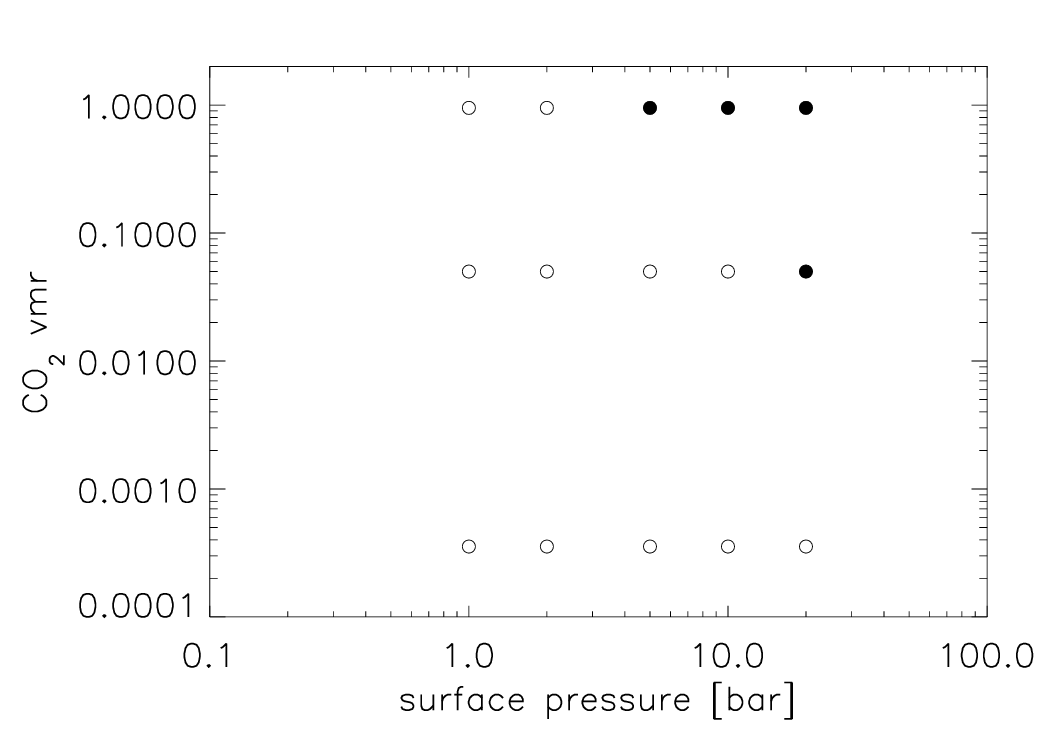}\\ \end{center}
  \caption{Assumed model atmospheres (Table \ref{listofruns}) for GL 581
  d: Habitable (filled circles) and uninhabitable (open circles) scenarios.}
  \label{habgrid}
\end{figure}

The results are summarized in Fig. \ref{habgrid} in the considered
parameter space, i.e. the surface pressure - CO$_2$ concentration
plane.

Yet with regard to the uncertainties in the radiative transfer
associated with CO$_2$ continuum absorption (see the sensitivity
study for surface temperature in Appendix \ref{sensstudy}), the
modeling results regarding the habitability must of course be
treated with caution.

Still, our results imply that, given plausible Venus- or early
Earth-like atmospheric compositions and taking into account
reasonable uncertainties in the radiative transfer formulations, GL
581 d can be classified as a habitable planet.

The study of \citet{bloh2007} concluded that GL 581 d represents a
habitable planet even for relatively low CO$_2$ partial pressures of
about 4-5 bar. Taking into account the more than 30 \% increase in
insolation for GL 581 d owing to the revision of orbital parameters
by \citet{mayor2009gliese}, the results of \citet{bloh2007} imply
that GL 581 d could indeed be habitable for even less dense CO$_2$
atmospheres. The findings of our study do not support these
implications of \citet{bloh2007} though.

This disagreement is partly owed to the parameterizations and
empirical criteria employed. For example, the parametrization of the
relation between CO$_2$ partial pressure, planetary albedo, and
surface temperature used by \citet{bloh2007} is based on work by
\citet{williams1997}, \citet{budyko1982} and
\citet{chamberlain1980}. These studies were done for Earth, orbiting
around the Sun, and thus did not account for different central stars
with a different spectral energy distribution or more massive
terrestrial planets. In contrast, the radiative-convective model
used here takes these effects into account.

The study of \citet{selsis2007gliese} concluded that a pure CO$_2$
atmosphere without additional greenhouse gases is unlikely to
provide habitable conditions on GL 581 d. Like in the study of
\citet{bloh2007}, they did this for the then known orbital
parameters of GL 581 d. \citet{selsis2007gliese} also provide a
discussion of the uncertainties of the limits of the outer HZ in
view of early Mars. From this discussion, GL 581 d could still be in
the HZ when CO$_2$ clouds or additional greenhouse gases are taken
into account. Note that, as already stated by
\citet{selsis2007gliese}, the use of parameterizations of the outer
boundary of the HZ provided by \citet{kasting1993} is uncertain for
low-mass stars below about 3,700 K effective temperature.

Based on the calculations by \citet{selsis2007gliese},
\citet{mayor2009gliese} concluded that GL 581 d is a habitable
planet, considering that the revised orbit means that it receives
more than 30 \% more stellar energy than previously thought. Because
according to our study the low CO$_2$ and most of the medium CO$_2$
scenarios are uninhabitable (see Figs. \ref{temperature_low} and
\ref{temperature_medium}), this disagrees with their general
implication. This can be motivated in more detail as follows: Given
that the equilibrium temperature of a planet only increases as the
fourth root of stellar energy input, the increase of received
stellar energy would only lead to an increase of about 10 K in
equilibrium temperature (from about 185 K to 195 K). Thus, only an
atmosphere providing a large greenhouse effect would be able to warm
the planet above 273 K, hence result in habitable conditions (see
for example Fig. \ref{temperature_high}).

The recent atmospheric modeling studies by \citet{wordsworth2010}
qualitatively agree with our simulations. \citet{wordsworth2010} use
a 1D radiative-convective model like ours. They also incorporate a
correlated-k approach for the radiative transfer, similar to what
what we did. However, they used a different parameterization of the
continuum absorption of CO$_2$ than the one used here (for a
sensitivity study of CO$_2$ continuum absorption, again see Appendix
\ref{sensstudy}). Their calculated surface temperatures for the
high-CO$_2$ cloud-free cases (set G3 in Table \ref{listofruns}) are
comparable with the results obtained here (about 310 K for a 10 bar
atmosphere, about 350 K for a 20 bar atmosphere).

\section{Conclusions}

\label{concl}

Detailed model calculations of possible atmospheres for the low-mass
extrasolar planet GL 581 d have been presented in this explorative
modeling study. Using an improved 1D radiative-convective climate
model, several key atmospheric parameters (e.g. surface pressure,
atmospheric composition) were varied to investigate their influence
on resulting surface conditions. The planetary scenarios
investigated here are consistent with assumptions made in the
literature regarding surface pressures and atmospheric compositions
of terrestrial planets.

GL 581 d is a potentially  habitable planet, because for massive
CO$_2$ atmospheres (5 or more bar surface pressure with CO$_2$
concentrations of 95 \%, 20 bar with 5 \% CO$_2$), the surface
temperatures exceeded 273 K, i.e. the freezing point of water. For
these massive CO$_2$ atmospheres, the surface temperatures could be
as high as 357 K.

The results of our high CO$_2$ scenarios are confirmed by the very
recent model results of \citet{wordsworth2010}. But our model
calculations imply that also atmospheres that are not
CO$_2$-dominated (i.e., 5\% vmr instead of 95\% vmr) could result in
habitable conditions for GL 581 d.

For atmospheric scenarios with less CO$_2$, the planet was found to
be uninhabitable in our calculations.

Nevertheless, GL 581 d is the first extrasolar (potentially
terrestrial) planet where habitable conditions are at least
conceivable within a reasonable range of surface pressures and
CO$_2$ concentrations.\newline

\begin{acknowledgements}

We thank the anonymous referee for his/her fast and detailed
comments, which helped improve the paper.

P.v.P. thanks J.W. Stock for useful discussions and comments on the
text.

This research has been supported by the Helmholtz Gemeinschaft
through the research alliance "Planetary Evolution and Life".

Pascal Hedelt acknowledges support from the European Research
Council (Starting Grant 209622: E$_3$ARTHs).

\end{acknowledgements}

\appendix

\section{Comparative studies used for model validations}
\label{comparison}

\subsection{Runs}

Several sets of comparison runs were performed to compare the model
results of the improved model with other published work and
benchmark calculations.

For all comparative runs, the central star was the Sun. The solar
input spectrum is based on the high-resolution spectrum provided by
\citet{gueymard2004} (e.g., \citealp{kitzmann2010}).

The first set of comparison runs tested the response of the model to
doubling/quadrupling the CO$_2$ content within the modern Earth
atmosphere to compare the results with investigations regarding the
anthropogenic greenhouse effect (runs W1-W8 in Table
\ref{greenhouse}).

The second set of comparison runs that tested the model sensitivity
to changes in CO$_2$ content was performed with a reduced solar
luminosity $S$ of 0.8 times the present value to compare to
published calculations of different models with the same assumptions
(runs S1-S6 in Table \ref{greenhouse}).

The simulation to which the runs from Table \ref{greenhouse} were
compared included an adjustment of the model surface albedo in a way
that modern Earth reference calculations yield a surface temperature
of 288 K. For the model used here, however, a reference surface
temperature of 284.5 K was chosen to account for the missing
greenhouse effect of methane, ozone, and nitrous oxide, which are
not included in the present model. This value of the surface
temperature is calculated with a modern Earth model
(\citealp{Grenf2007pss}, when using the RRTM radiative transfer
scheme developed by \citealp{Mlawer1997}) upon excluding the
greenhouse effect provided by methane, ozone, and nitrous oxide. The
model surface albedo is 0.24, which is slightly higher than the
value of 0.21 from \citet{vparis2008}, but still compatible with
surface albedos of other cloud-free models in the literature (e.g.,
\citealp{goldblatt2009faintyoungsun}, \citealp{haqq2009}, both
studies using 0.23 as their surface albedo value). Note that the
actual global value for Earth is approximately 0.13
(\citealp{kitzmann2010}, \citealp{rossow1999}), as stated above.

\begin{table}[H]
  \caption{Runs performed to test the sensitivity
  of the model to variations of CO$_2$ content
  }\label{greenhouse}
\begin{center}
\begin{tabular}{clcl}
 \hline
   Run  &  $S$ & $p_{\mathrm{N_2}}$ [bar]  & $p_{\mathrm{CO_2}}$ [bar] \\
  \hline
   W1   & 1    & 0.77                      &  10$^{-4}$ \\
   W2   & 1    &0.77                       &  3.55 $\cdot$ 10$^{-4}$\\
   W3   & 1    &0.77                       &  7.1 $\cdot$ 10$^{-4}$\\
   W4   & 1    &0.77                       &  10$^{-3}$\\
   W5   & 1    &0.77                       &  1.42 $\cdot$ 10$^{-3}$\\
   W6   & 1    &0.77                       &  2.84 $\cdot$ 10$^{-3}$\\
   W7   & 1    &0.77                       &  5.68 $\cdot$ 10$^{-3}$\\
   W8   & 1    &0.77                       &  10$^{-2}$\\
  \hline
   S1   & 0.8  & 0.8                       & 10$^{-3}$ \\
   S2   & 0.8  & 0.8                       & 10$^{-2}$\\
   S3   & 0.8  & 0.8                       & 1.5 $\cdot$ 10$^{-2}$\\
   S4   & 0.8  & 0.8                       & 10$^{-1}$\\
   S5   & 0.8  & 0.8                       & 1.\\
   S6   & 0.8  & 0.8                       & 10.0\\

\end{tabular}
\end{center}

\end{table}

The third set of comparison runs followed a proposed evolutionary
sequence of the Earth's atmosphere based on \citet{hart1978} in
terms of time $t_b$ before present (runs H1-H8 in Table \ref{hart}).
To this end, the model surface albedo was adjusted to yield a
surface temperature of 288 K for present luminosity to be consistent
with model studies using this sequence, which also introduced this
particular adjustment (\citealp{kasting1986}, \citealp{kiehl1987}).
The resulting surface albedo was 0.185, in contrast to the assumed
0.24 for all other runs. The assumed N$_2$ partial pressure was 770
mb.

\begin{table}[H]
  \caption{Runs for the evolution sequence of \citet{hart1978}
  }\label{hart}
\begin{center}
\begin{tabular}{clcc}
 \hline
   Run  &  $t_b$ [Gy]& $S$  & $c_{\mathrm{CO_2}}$ [vmr] \\
  \hline
   H1   & 0.0        & 1.0  & 3.3 10$^{-4}$ \\
   H2   & 0.5        &0.972 & 3.3 10$^{-4}$\\
   H3   & 1.0        &0.944 & 6.5 10$^{-4}$\\
   H4   & 1.5        & 0.917& 2.9 10$^{-3}$ \\
   H5   & 2.0        &0.889 & 8.6 10$^{-3}$\\
   H6   & 2.5        &0.861 & 1.8 10$^{-2}$\\
   H7   & 3.0        & 0.833&  3.3 10$^{-2}$\\
   H8   & 4.25       & 0.764& 0.31 \\
\end{tabular}
\end{center}

\end{table}

\subsection{Results from comparative studies}

\begin{figure}[H]
 \resizebox{\hsize}{!}{  \includegraphics[width=220pt]{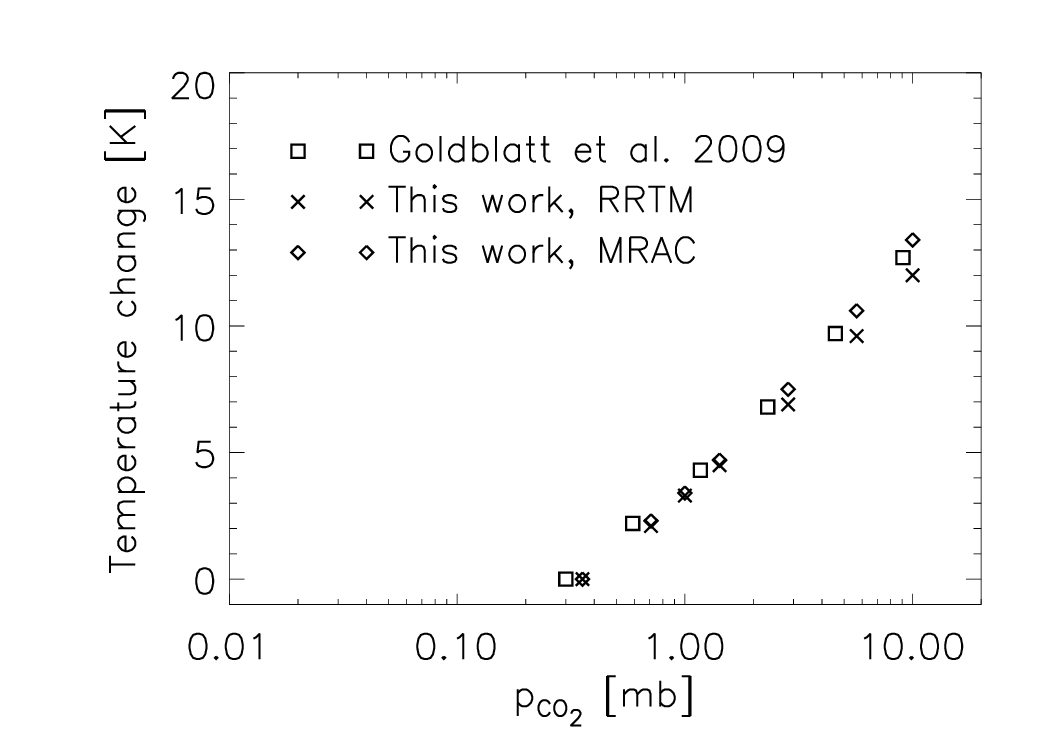}}\\
  \caption[Response of surface temperature to increases in CO$_2$]{Response of surface temperature to increases in CO$_2$:
  Temperature change with respect to run W1 . }\label{ghe_comp}
\end{figure}

Figure \ref{ghe_comp} summarizes the results of the runs of set W
(Table \ref{greenhouse}), i.e. tests for the assumed anthropogenic
greenhouse effect. Shown is the increase $\Delta T$ of surface
temperature with respect to run W1, which is a measure of the
strength of the greenhouse effect caused by CO$_2$.

Figure \ref{ghe_comp} also shows values for the same numerical
experiment performed with two other models. The first model is
actually a version of the model used here, but it uses RRTM
\citep{Mlawer1997} instead of MRAC (see model description above) in
the IR radiative transfer. The second model is from
\citet{goldblatt2009} (also used in
\citealp{goldblatt2009faintyoungsun}), which is also a 1D
radiative-convective model. It differs from the model used here in
that the correlated-k radiative transfer in the IR is based on
Hitran 1992. Furthermore, the radiative transfer uses a 2-stream
method instead of the diffusivity approximation used by MRAC. The
main difference though is the numerical scheme employed to reach the
steady state atmosphere. \citet{goldblatt2009} use a Newton-Raphson
method, not a time-stepping algorithm. Figure \ref{ghe_comp} shows
that the agreement between the three models is indeed very good.

\begin{figure}[H]
  \resizebox{\hsize}{!}{\includegraphics[width=320pt]{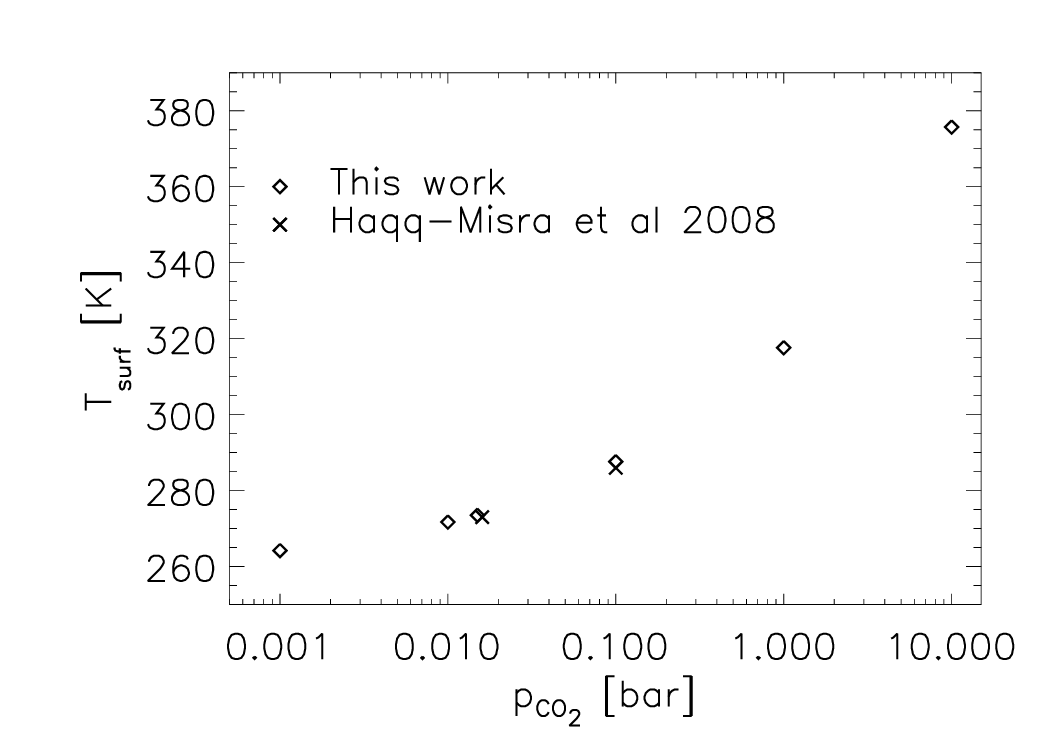}}\\
  \caption[Response of surface temperature to increasing CO$_2$ partial pressures at a
   reduced solar constant]{Response of surface temperature to increasing CO$_2$ partial pressures at a
   reduced solar constant (0.8 times present-day value; set S from Table \ref{greenhouse})}\label{kasting_ackerman}
\end{figure}

Figure \ref{kasting_ackerman} compares the calculations for set S
from Table \ref{greenhouse} with calculations done by
\citet{haqq2009} (similar to runs S3 and S4). They used an updated
version (e.g., new absorption coefficients) of the 1D
radiative-convective model of \citet{kasting1986}. The difference
between their model and ours is that the IR radiative transfer is
performed by exponential-sum fitting of transmission instead of
correlated-k. Figure \ref{kasting_ackerman} implies that the results
from \citet{haqq2009} and this work for the two specific runs S3 and
S4 agree very well.

\begin{figure}[H]
  \resizebox{\hsize}{!}{\includegraphics[width=320pt]{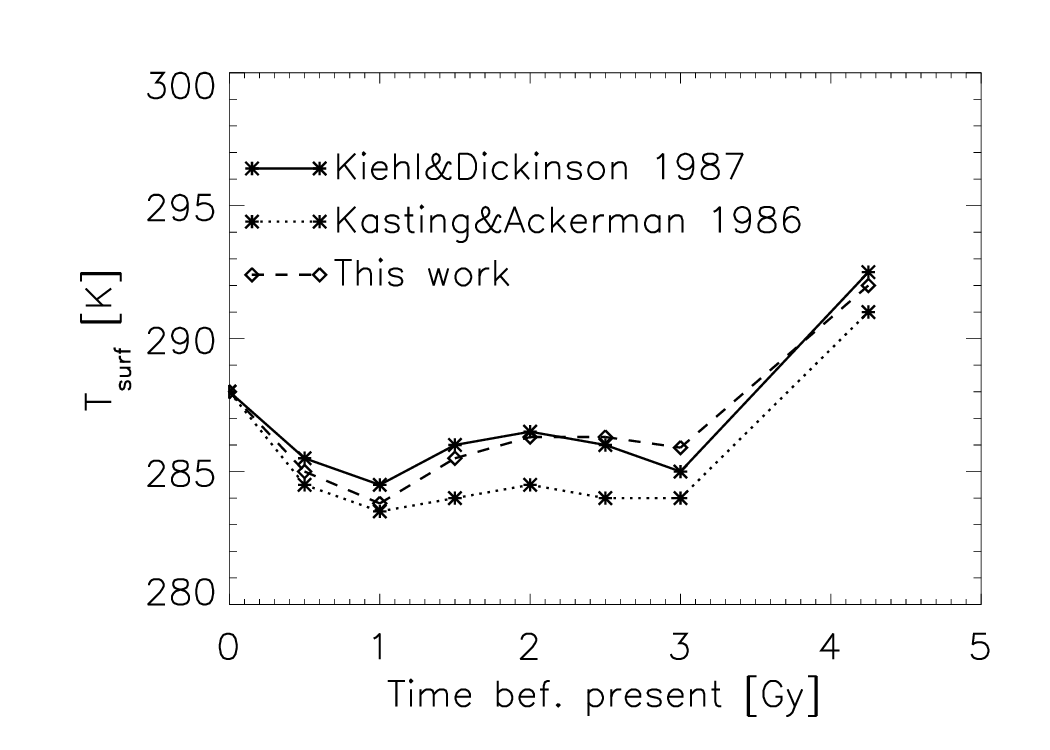}}\\
  \caption[Surface temperatures calculated for the evolutionary sequence of \citet{hart1978}]{Surface temperatures calculated for the evolutionary sequence of \citet{hart1978}, set H from Table \ref{hart}.}
  \label{hartcomp}
\end{figure}

The results of the model calculations for set H from Table
\ref{hart} were compared with the model results of \citet{kiehl1987}
and \citet{kasting1986} (see Figs. 7 and 8 in \citealp{kiehl1987};
Fig. 2 in \citealp{kasting1989}) and are shown in Fig.
\ref{hartcomp}. All models included in this test are 1D
radiative-convective models.

The model by \citet{kasting1986} uses band models and exponential
sums in the radiative transfer instead of the correlated-k. Also,
they calculated the relative humidity in the atmosphere by a
slightly different approach than the one used by us.

The study by \citet{kiehl1987} uses a relatively high-resolution
radiative transfer code (500 intervals in the thermal region) based
on band models. A second difference is the convective adjustment
scheme that is not based on a simple formulation of the adiabatic
lapse rate, the approach used by \citet{kasting1986} and here.

Calculated surface temperatures are within 2 K of the results of the
two other studies. These differences are quite small given that the
models differ with respect to e.g., radiative transfer and the
treatment of convection, as stated above.

\subsection{Conclusions from comparative studies}

The purpose of this section was to test the updated and improved
model against other published model results. It was shown that the
model agrees well with recent models of \citet{vparis2008} and
\citet{haqq2009} as well as earlier model calculations
(\citealp{kiehl1987}; \citealp{kasting1986}). Furthermore, by
comparing with benchmark studies regarding the anthropogenic
greenhouse effect, the model was found to agree also very well with
other model results (e.g., \citealp{goldblatt2009}).

\section{Sensitivity studies for the effect of radiative transfer approximations on surface temperature}

\label{sensstudy}

In order to assess the sensitivity of the surface temperatures
calculated for the atmospheric scenarios of Table \ref{listofruns}
to details in the radiative transfer (such as continuum absorption
or line mixing), an additional sensitivity study was performed.
Based on this sensitivity study, the results regarding surface
temperatures and habitability can be interpreted with more
confidence. Also, it serves as an illustration of the uncertainties
associated with atmospheric modeling in general.

The sensitivity runs have been performed for the simulations with
medium and high CO$_2$ concentrations (sets G2-G3 in Table
\ref{listofruns}), because these runs resulted in habitable surface
conditions.

\subsection{Performed runs}

Both H$_2$O and CO$_2$ show significant collision-induced continuum
absorption in the mid- to far-infrared. Especially the formulations
of the foreign continua of H$_2$O and CO$_2$ could be relatively
uncertain, because they are calculated for N$_2$-O$_2$ background
atmospheres (see, e.g. \citealp{halevy2009}).

The first set of sensitivity runs was performed in order to assess
the influence of the H$_2$O continuum on the surface temperature
(runs CH1-CH4 in Table \ref{sensrunstab}). These tests were done for
the 5, 10, and 20 bar atmosphere of set G3 as well as the 20 bar run
from set G2 (Table \ref{listofruns}) by multiplying the H$_2$O
continuum by arbitrary factors $f_{\mathrm{H_2O}}$ of 0.2 and 0.5.

The second set of sensitivity runs was performed to assess the
influence of the CO$_2$ self-continuum on the surface temperature
(runs CS1-CS4 in Table \ref{sensrunstab}). The CO$_2$ self-continuum
was multiplied by arbitrary factors $f_{\mathrm{CO_2}}$ of 0.2 and
0.5, respectively. To test the influence of CO$_2$ foreign-continuum
absorption, the continuum was removed for the same runs, i.e.
multiplied by 0 (runs CF1-CF4 in Table \ref{sensrunstab}). These
tests were performed for the same runs as the H$_2$O sensitivity
runs.

\begin{table}[H]
  \caption[Sensitivity runs performed for GL 581 d]
  {Sensitivity runs performed for GL 581 d ($f_{\mathrm{H_2O}}$ H$_2$O continuum;
   $f_{\mathrm{CO_2}}$ CO$_2$ continuum; $f_{\mathrm{LM}}$ line mixing factor)}
  \label{sensrunstab}
\resizebox{\hsize}{!}{\begin{tabular}{ccccc}
 \hline
    Set         & Control Run          &  $f_{\mathrm{H_2O}}$      & $f_{\mathrm{CO_2}}$     & $f_{\mathrm{LM}}$\\
  \hline
    CH1         & G3 20 bar            & 0.2,0.5              & 1                       &1\\
    CH2         & G3 10 bar            & 0.2,0.5              & 1                       &1\\
    CH3         & G3 5 bar             & 0.2,0.5              & 1                       &1\\
    CH4         & G2 20 bar            & 0.2,0.5              & 1                       &1\\
  \hline
    CS1         & G3 20 bar            & 1                         & 0.2,0.5             &1\\
    CS2         & G3 10 bar            & 1                         & 0.2,0.5             &1\\
    CS3         & G3 5 bar             & 1                         & 0.2,0.5             &1\\
    CS4         & G2 20 bar            & 1                         & 0.2,0.5             &1\\
  \hline
    CF1         & G3 20 bar            & 1                         & 0             &1\\
    Cf2         & G3 10 bar            & 1                         & 0             &1\\
    CF3         & G3 5 bar             & 1                         & 0             &1\\
    CF4         & G2 20 bar            & 1                         & 0             &1\\
  \hline
    LM1         & G3 20 bar            & 1                         & 1                       &0.2,0.5\\
    LM2         & G3 10 bar            & 1                         & 1                       &0.2,0.5\\
    LM3         & G3 5 bar             & 1                         & 1                       &0.2,0.5\\
    LM4         & G2 20 bar            & 1                         & 1                       &0.2,0.5\\
\end{tabular}}
\end{table}

An additional challenge in the radiative transfer arises from
so-called line-mixing. Assuming a Voigt line profile (i.e. a
convolution of Lorentz and Gauss profiles) is no longer justified.
Comparisons of computer simulations with experimental data by
\citet{rodrigues1999} have shown that including line-mixing into the
calculations can result in a decrease of absorption coefficients by
up to a factor of 2. Line-mixing parameters for CO$_2$ are included
in the HITRAN database \citep{rothman2005}, but not in the HiTemp
database \citep{rothman1995} used for the absorption coefficients in
the model. Hence, a Voigt profile was assumed for all lines at all
pressures during the calculations of the absorption coefficients
\citep{vparis2008}. In order to test the sensitivity of the results
to line-mixing, a third set of sensitivity runs was performed (runs
LM1-LM4 in Table \ref{sensrunstab}). Here, the optical depth in the
main atmospheric band of CO$_2$ (15 $\mu$m) was arbitrarily
decreased in the lower troposphere (i.e. at pressures higher than
100 mb) by factors $f_{\mathrm{LM}}$ of 0.2 and 0.5, respectively.
Again, these tests were performed for the same runs as the H$_2$O
sensitivity runs.

Table \ref{sensrunstab} summarizes the sensitivity runs.

\subsection{Effect of H$_2$O continuum}

The runs of sets CH1-CH4 from Table \ref{sensrunstab} were done in
order to test the influence of the H$_2$O continuum on calculated
surface temperatures. The obtained results did not differ
significantly from the control runs. For a decrease of H$_2$O
continuum of a factor of 5, the corresponding decrease of surface
temperature was less than 2 K in each case.

Thus, these results indicate that the formulation of the H$_2$O
continuum is not critical for the calculations presented here.

\subsection{Effect of CO$_2$ continuum}

Sensitivity tests were performed to assess the influence of the
CO$_2$ continuum on surface temperature (sets CS1-CS4 and CF1-CF4
from Table \ref{sensrunstab}). For the CO$_2$ self-continuum, the
effect was relatively large. The strong mid-IR band around 7 $\mu$m
dominates absorption and is the main opacity source in this spectral
region. Also, the far-IR bands show significant contribution to the
overall absorption. Hence, the effect of a decrease of the CO$_2$
continuum can be expected to be very strong.

For set CS1, the surface temperature decreased from 357 K (control
run) to 346 K and 329 K on decreasing the continuum absorption by a
factor of 2 and 5, respectively.

Table \ref{cceffect} summarizes the effect on surface temperatures.

\begin{table}[H]
  \caption[Surface temperatures (in K) for the sets CS1-CS4]
  {Surface temperatures (in K) for the sets CS1-CS4 from Table \ref{sensrunstab}.}
  \label{cceffect}
\begin{center}
\resizebox{\hsize}{!}{\begin{tabular}{cccccc}
 \hline
    Set   & Control Run   &  $f_{\mathrm{CO_2}}$=0.2      & $f_{\mathrm{CO_2}}$=0.5     \\
  \hline
   CS1    & 357           &   329                         & 346                        \\
   CS2    & 322           &   292                         & 311                        \\
   CS3    & 287           &   257                         & 274                        \\
   CS4    & 313           &   282                         & 300                        \\
  \end{tabular}}
\end{center}

\end{table}

From the results in Table \ref{cceffect}, one infers that the
uncertainties in the CO$_2$ self continuum opacity will not alter
the principle conclusions on surface habitability provided that they
do not exceed a factor of 2-5. The results of the sensitivity tests
imply, however, that more detailed modeling and measurements of the
CO$_2$ self-continuum absorption are needed in the future.

Upon removing the CO$_2$ foreign continuum (sets CF1-CF4 in Table
\ref{sensrunstab}), surface temperatures decrease by 1-3 K for the
high CO$_2$ runs (CF1-CF3). For the 20 bar medium CO$_2$ run,
however, excluding the CO$_2$ foreign continuum decreased the
surface temperature by 33 K to 280 K. This implies that the foreign
continuum is an important opacity source and should be included in
all future simulations.

\subsection{Effect of line mixing}

The effect of line mixing on surface temperatures was investigated
with the sets LM1-LM4 from Table \ref{sensrunstab}.

For all cases, the surface temperatures were almost unaffected (less
than 1 K decrease). These effects are important for interpreting
measurements on Mars or Venus, but are not likely to significantly
affect results regarding exoplanets, where only first-order
estimates can be done so far.

\subsection{Summary of sensitivity studies}

The purpose of this section was to investigate the influence of the
radiative transfer approximations on the calculated surface
temperatures.

It was shown that the model results do depend to a certain extent on
the details of the radiative transfer. However, only the effect of
the CO$_2$ continuum is likely to be important for our study.

\bibliographystyle{aa}
\bibliography{literatur_phd}

\end{document}